\documentclass[twocolumn]{aa}
\usepackage{graphicx}
\usepackage{txfonts}
\begin{document}
\title{An investigation of \ion{Fe}{xv} emission lines in solar flare spectra}

\author{F. P. Keenan
          \inst{1}
          \and
          K. M. Aggarwal
          \inst{1}
          \and
          D. S. Bloomfield
          \inst{1}
          \and
          A. Z. Msezane
          \inst{2}     
          \and
          K. G. Widing\inst{3}
          }

\offprints{F. P. Keenan}

\institute{Department of Physics and Astronomy, Queen's University Belfast, Belfast BT7 1NN, 
Northern Ireland, UK
\\
\email{F.Keenan@qub.ac.uk}
\and
Center for Theoretical Studies of Physical Systems,
Clark Atlanta University, Atlanta, GA 30304, USA
\and
Computational Physics, Inc., 8401 Braddock Road, Springfield, VA 22151, USA and 
Naval Research Laboratory, Washington, DC 20375-5320, USA
}

\date{Received ; accepted }

\abstract{
Previously, large discrepancies have been found between theory and observation for 
\ion{Fe}{xv} emission line ratios in solar flare spectra covering the 224--327~\AA\ 
wavelength range, obtained by the Naval Research Laboratory's S082A instrument on board 
{\em Skylab}. These discrepancies have been attributed to either errors in the adopted
atomic data or the presence of additional atomic processes not included in the modelling, 
such as fluorescence. However our analysis of these plus other S082A flare observations 
(the latter containing \ion{Fe}{xv} transitions between 321--482~\AA),
performed using the most recent \ion{Fe}{xv} atomic physics calculations in conjunction 
with a {\sc chianti} synthetic flare spectrum, indicate that blending of the 
lines is primarily responsible for the discrepancies. As a result, most \ion{Fe}{xv} 
lines cannot be employed as electron density diagnostics for solar flares, at least 
at the spectral resolution of S082A and similar instruments (i.e. $\sim$0.1~\AA).
An exception is the intensity ratio I(3s3p $^{3}$P$_{2}$--3p$^{2}$ 
$^{3}$P$_{1}$)/I(3s3p $^{3}$P$_{2}$--3p$^{2}$ $^{1}$D$_{2}$) = I(321.8~\AA)/I(327.0~\AA),
which appears to provide good estimates of the electron density at this spectral
resolution.

}

\maketitle

%

\section{Introduction}

Emission lines of \ion{Fe}{xv} are prominent features of the solar extreme ultraviolet (EUV)
spectrum, with numerous transitions present in the $\sim$220--500~\AA\ wavelength interval
(see, for example, Dere 1978; Thomas \& Neupert 1994). It has long been known that 
these lines provide useful electron density diagnostics for the emitting plasma
(Bely \& Blaha 1968), but to date there have been relatively few detailed analyses of 
the \ion{Fe}{xv} solar EUV spectrum. Probably the most comprehensive have been those of
Young et al. (1998) and Keenan et al. (2005),  both of which employed high 
resolution solar active region spectra obtained by the Solar EUV Research Telescope and 
Spectrograph (SERTS), and generally found good agreement between theory and observation.

However the situation for solar flares is very different. The most detailed study 
of \ion{Fe}{xv} flare lines was undertaken by Dufton et al. (1990), using spectra from
the S082A instrument on board {\em Skylab}. These authors found major discrepancies 
between theory and observation, which they attributed to possible errors in the 
adopted atomic data, especially for higher-lying levels, which would be more populated 
in flares than in active regions due to the higher electron densities. 
Alternatively, Dufton et al. suggested that
some additional atomic process could be in operation, which may be more 
important in flares than in active regions, such as fluorescence. 

In this paper we use the most recent atomic physics calculations for \ion{Fe}{xv}, in 
conjunction with a synthetic flare spectrum generated with the {\sc chianti}
database (Dere et al. 1997; Young et al. 2003), to investigate if the 
discrepancies found between
theory and observation by Dufton et al. (1990)
may be resolved. 
In addition, we perform an analysis of other S082A flare spectra which contain \ion{Fe}{xv}
emission lines not considered by these authors.

\section{Observational data}

The solar flare observational data analysed in the present paper 
have been obtained 
with the Naval Research Laboratory's slitless
spectrograph (S082A) on board {\em Skylab}.
This instrument operated in the 171--630~\AA\
wavelength range in two sections
(171--350~\AA\ and 300--630~\AA), and produced dispersed
images of the Sun
on photographic film with a spatial resolution of 2\arcsec\
and a maximum
spectral resolution of $\sim$0.1~\AA\ (FWHM).
Details of the S082A spectrograph and reduction procedures
can be found in Parker et al. (1976) and Tousey et al. (1977),
and the absolute calibration curves in Dere \& Mango (1979)
and Dufton et al. (1983) for
wavelengths $>$260~\AA\
and $\leq$260~\AA, respectively. 
We note that the sensitivity curve for S082A 
is based on the calibration using a synchrotron source at the
National Bureau of Standards of an instrument similar to, but smaller than,
the {\em Skylab} spectrograph. 
The absolute values of line intensities from the S082A spectra are only accurate
to probably a factor of 2 (Dere 1982). However the relative calibration 
of the instrument is more secure. As a result, line intensity ratios
can generally be determined to an accuracy of between $\pm$30\%\ 
to $\pm$40\%\ (Dere 1982; Keenan et al. 1984;
Widing et al. 1986). An exception is when the lines are close to the
short or long wavelength cutoffs of the plates, where aberration leads to significant
image broadening. For example, at 320~\AA\ the spatial resolution of S082A is
about a factor of 1.4 lower than the maximum value of 
2\arcsec\ (Tousey et al. 1977). In principle, line intensities 
at these wavelengths may be corrected for the presence of aberration, 
by multiplying the measured values by an aberration factor. However this 
is generally
more successful for stronger emission lines.

Dufton et al. (1990) have presented a comprehensive set of \ion{Fe}{xv}
line ratios for several solar flares, measured from S082A short wavelength
(171--350~\AA) plates. Details of these observations may be found in that paper. 
However, briefly Dufton et al. measured intensities for 2 groups of 
\ion{Fe}{xv} transitions, one covering $\sim$224--244~\AA\ containing 
3s3p $^{3}$P$_{J}$--3s3d $^{3}$D$_{J^\prime}$ lines plus the 
3s3p $^{1}$P$_{1}$--3s3d $^{1}$D$_{2}$ feature at 243.8~\AA.
The other group spanned $\sim$292--327~\AA, and observed the
3s3p $^{3}$P$_{J}$--3p$^{2}$ $^{1}$D$_{2}$ and 
3s3p $^{3}$P$_{J}$--3p$^{2}$ $^{3}$P$_{J^\prime}$ lines.
However the second set of data are less reliable due to contamination by an image of the solar
disk in the strong \ion{He}{ii} 303.8~\AA\ line. Although the intense \ion{Fe}{xv}
3s$^{2}$ $^{1}$S$_{0}$--3s3p $^{1}$P$_{1}$ resonance line at 284.2~\AA\ also lies within the
S082A wavelength coverage, it is normally saturated
on the photographic emulsion and hence was not measured by Dufton et al.

As noted above, Dufton et al. (1990) only considered short wavelength 
plates from the S082A instrument. However several \ion{Fe}{xv} transitions are also detected
on long wavelength (300--630~\AA) plates. These
are listed in Table 1,  
where the identifications are from Dere (1978).
We have measured intensities
for these lines relative to the 
3s$^{2}$ $^{1}$S$_{0}$--3s3p $^{3}$P$_{1}$ transition at 417.3~\AA\ (the strongest 
\ion{Fe}{xv} line predicted in the 300--630~\AA\ region) in several solar flares. These
are the events of
1973 June 15 at 14:27:20 UT (discussed in detail by Widing \& Cheng 1974; Widing \& Dere 1977), 
1973 August 9 at 
15:55:05 UT (Dere \& Cook 1979; Dere et al. 1979), and
1973 December 17 at
00:48:49 UT (Widing \& Spicer 1980; Widing \& Cook 1987). 
In Table 1 
we list the measured intensities of the 417.3~\AA\ line in the solar flares; values 
for the other transitions may be inferred from the observed line ratios, also given
in the table. As noted previously, the S082A relative calibration 
is better determined than the absolute one. Hence the errors in the
measured line ratios are lower than what would be expected
on the basis of uncertainties
in the absolute intensities.

%
%
%
\begin{table}
\caption{\ion{Fe}{xv} line intensity ratios in S082A flare spectra}       
\label{table:1}      
\begin{tabular}{l c c c}      
\hline               
Flare & $\lambda$$_{1}$ (\AA) & $\lambda$$_{2}$ (\AA) & R = I($\lambda$$_{1}$)/I($\lambda$$_{2}$) 
\\  
\hline    
1973 June 15 & 321.8$^{\mathrm{a}}$ & 417.3$^{\mathrm{b}}$ & 0.26$\pm$0.08 
\\ 
 & 327.0 & 417.3    & 0.41$\pm$0.12 
\\
 & 481.5$^{\mathrm{d}}$ & 417.3    & 0.18$\pm$0.05 
\\
1973 August 9 & 321.8 & 417.3$^{\mathrm{c}}$ & 1.0$\pm$0.3 
\\
  & 327.0 & 417.3  & 1.3$\pm$0.4 
\\
  & 481.5 & 417.3 & 0.26$\pm$0.08
\\
1973 December 17 & 321.8 & 417.3$^{\mathrm{d}}$ & 0.37$\pm$0.11
\\
  & 327.0 & 417.3 & 0.38$\pm$0.11
\\
  & 481.5 & 417.3 & 0.23$\pm$0.07 
\\
\hline
\end{tabular}

$^{\mathrm{a}}$ The wavelengths listed correspond to the \ion{Fe}{xv}
transitions: 3s3p $^{3}$P$_{2}$--3p$^{2}$ $^{3}$P$_{1}$ (321.8~\AA);
3s$^{2}$ $^{1}$S$_{0}$--3s3p $^{3}$P$_{1}$ (417.3~\AA);
3s3p $^{3}$P$_{2}$--3p$^{2}$ $^{1}$D$_{2}$ (327.0~\AA);
3s3p $^{1}$P$_{1}$--3p$^{2}$ $^{1}$D$_{2}$ (481.5~\AA).

$^{\mathrm{b}}$ I(417.3~\AA) = (2.5$\pm$1.3)$\times$10$^{23}$ erg s$^{-1}$.

$^{\mathrm{c}}$ I(417.3~\AA) = (1.6$\pm$0.8)$\times$10$^{23}$ erg s$^{-1}$.

$^{\mathrm{d}}$ I(417.3~\AA) = (5.0$\pm$2.5)$\times$10$^{22}$ erg s$^{-1}$.
\end{table}

\section{Theoretical line ratios}

The model ion adopted for \ion{Fe}{xv}
has been discussed in detail by Keenan et al. (2005). Briefly,
the energetically lowest 53 fine-structure levels belonging to
the 3s$^{2}$, 3s3p, 3p$^{2}$, 3s3d, 3p3d,
3d$^{2}$, 3s4s, 3s4p, 3s4d, 3p4s and 3s4f configurations were included
in the calculations.
Only collisional excitation and de-excitation by electrons and protons (the latter 
in the case of transitions among the 3s3p $^{3}$P levels) and spontaneous radiative
de-excitation processes were considered. Sources of the atomic data 
were Deb et al. (1999), Aggarwal et al. (2003) and Landman \& Brown (1979)
for A-values, electron and proton impact excitation rates, respectively.
These papers either contain full listings of atomic data, 
or links to websites where the results may be freely accessed,
hence allowing readers to reproduce our calculations if desired.
All emission lines in our calculations were assumed to 
be optically thin,  which is a valid assumption, as although the 
3s$^{2}$ $^{1}$S$_{0}$--3s3p $^{1}$P$_{1}$ resonance line may be optically thick,
this does not affect predictions of intensity ratios involving other \ion{Fe}{xv}
transitions (Kastner \& Bhatia 2001). 
Further details of the line ratio calculations
may be found in Keenan et al., where it is noted that the accuracy 
of the results should be better than $\pm$20\%.

We have generated theoretical \ion{Fe}{xv} line ratios
for a grid of electron temperature (T$_{\mathrm{e}}$) and density (N$_{\mathrm{e}}$)
values, with T$_{\mathrm{e}}$ = 10$^{5.9}$--10$^{6.9}$~K in steps of 0.1~dex,
and N$_{\mathrm{e}}$ = 10$^{8}$--10$^{13}$~cm$^{-3}$ also in steps of 0.1~dex. 
These are far too extensive to reproduce here, as with 53 fine-structure levels in our 
calculations we have intensities for 1378 transitions at each of the 
561 possible (T$_{\mathrm{e}}$, N$_{\mathrm{e}}$) combinations considered. 
However, results 
involving any line pair, in either photon or energy units, are freely available from
one of the authors (FPK) by email on request.

The present line ratio calculations are quite similar to those generated 
by Dufton et al. (1990). This is illustrated in Figs. 1 and 2, where we plot 
the intensity ratios I(224.8~\AA)/I(233.9~\AA) and I(243.8~\AA)/I(233.9~\AA), respectively,
at the temperature of maximum \ion{Fe}{xv}
fractional abundance in ionization equilibrium, T$_{\mathrm{e}}$ = 
10$^{6.3}$~K (Mazzotta et al.
1998). An inspection of the figures reveals that the new I(224.8~\AA)/I(233.9~\AA)
ratios differ by negligible amounts from those of Dufton et al. 
For the I(243.8~\AA)/I(233.9~\AA)
ratios, the discrepancies are larger, but are still only about 30\%\ over the electron density
interval N$_{\mathrm{e}}$ = 10$^{8}$--10$^{10}$~cm$^{-3}$, decreasing to less than 15\%\
for N$_{\mathrm{e}}$ $\geq$ 10$^{11}$~cm$^{-3}$.

%
%
\begin{figure}
\centering
\includegraphics[width=6cm,angle=90]{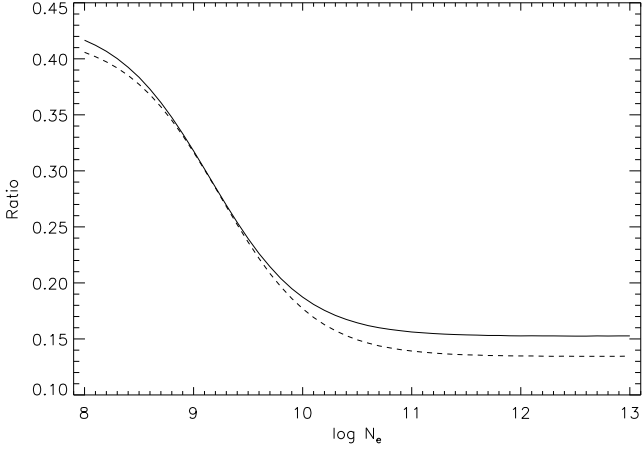}
\caption{The theoretical \ion{Fe}{xv}
emission line intensity ratio 
I(3s3p $^{3}$P$_{0}$--3s3d $^{3}$D$_{1}$)/I(3s3p $^{3}$P$_{2}$--3s3d $^{3}$D$_{3}$) 
= I(224.8~\AA)/I(233.9~\AA),
where I is in energy units,
plotted as a function of logarithmic electron density
(N$_{\mathrm{e}}$ in cm$^{-3}$) at an electron temperature
of T$_{\mathrm{e}}$ = 10$^{6.3}$~K, with:
solid line --- present calculations; dashed line --- results of Dufton et al. (1990).
}
\label{}
\end{figure}
%
%

%
%
\begin{figure}
\centering
\includegraphics[width=6cm,angle=90]{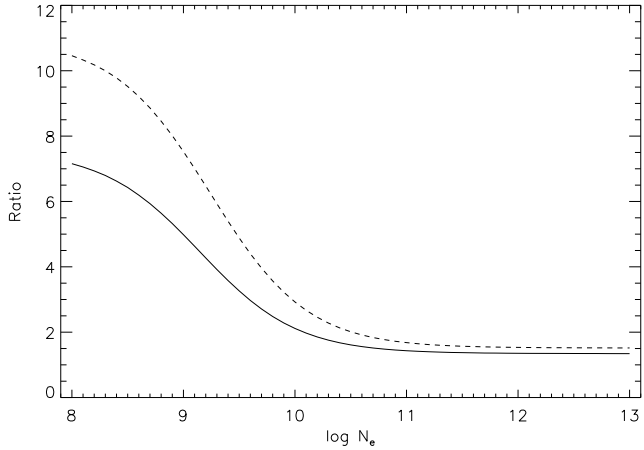}
\caption{The theoretical \ion{Fe}{xv}
emission line intensity ratio 
I(3s3p $^{1}$P$_{1}$--3s3d $^{1}$D$_{2}$)/I(3s3p $^{3}$P$_{2}$--3s3d $^{3}$D$_{3}$) 
= I(243.8~\AA)/I(233.9~\AA),
where I is in energy units,
plotted as a function of logarithmic electron density
(N$_{\mathrm{e}}$ in cm$^{-3}$) at an electron temperature
of T$_{\mathrm{e}}$ = 10$^{6.3}$~K, with: solid line --- present calculations; 
dashed line --- results of Dufton et al. (1990).
}
\label{}
\end{figure}

In the case of the longer wavelength lines in the S082A spectra not considered by Dufton 
et al. (1990), we note that several of their intensity ratios 
are predicted to be electron density sensitive. For example, in Fig. 3 we plot the
I(321.8~\AA)/I(417.3~\AA) ratio at 3 electron temperatures, namely that of maximum 
fractional abundance for \ion{Fe}{xv} in ionization equilibrium, T$_{\mathrm{e}}$
= 10$^{6.3}$~K, plus $\pm$0.3 dex about this value. 
An inspection of the figure reveals that the ratio varies by a factor of
about 6 over the density interval N$_{\mathrm{e}}$ = 10$^{8}$--10$^{11}$~cm$^{-3}$,
while being relatively insensitive to the adopted electron temperature.
Hence in principle it should provide 
an excellent density diagnostic.

%
%
\begin{figure}
\centering
\includegraphics[width=6cm,angle=90]{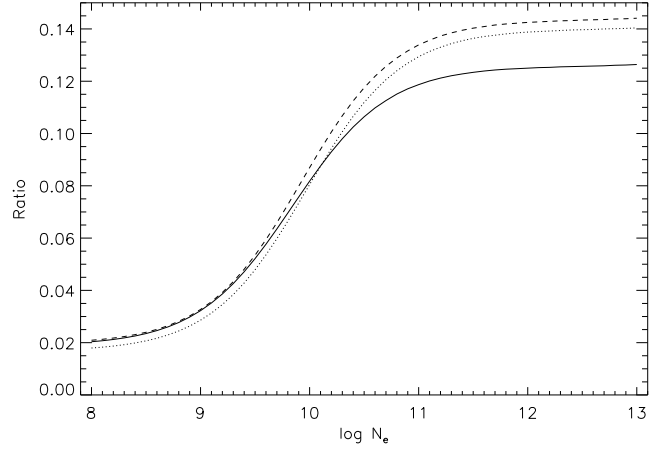}
\caption{Present calculations of the \ion{Fe}{xv}
emission line intensity ratio I(3s3p $^{3}$P$_{2}$--3p$^{2}$ $^{3}$P$_{1}$)/I(3s$^{2}$
$^{1}$S$_{0}$--3s3p $^{3}$P$_{1}$) = 
I(321.8~\AA)/I(417.3~\AA),
where I is in energy units,
plotted as a function of logarithmic electron density
(N$_{\mathrm{e}}$ in cm$^{-3}$) at electron temperatures
of: T$_{\mathrm{e}}$ = 10$^{6.0}$~K (solid line); T$_{\mathrm{e}}$
= 10$^{6.3}$~K (dashed line); T$_{\mathrm{e}}$
= 10$^{6.6}$~K (dotted line).
}
\label{}
\end{figure}
%
%

\section{Results and discussion}

As noted in Section 1, Dufton et al. (1990) found large discrepancies between theory and 
observation for \ion{Fe}{xv} line ratios measured from solar flare spectra obtained 
with the S082A instrument on board {\em Skylab}. Specifically, they noted that:
\begin{itemize}
\item
the intensities of the 
3s3p $^{3}$P$_{J}$--3s3d $^{3}$D$_{J^\prime}$ transitions between
$\sim$224--235~\AA, and that of the  
3s3p $^{1}$P$_{1}$--3s3d $^{1}$D$_{2}$ line at 243.8~\AA,
when ratioed against the 
3s3p $^{3}$P$_{2}$--3s3d $^{3}$D$_{3}$ feature at 233.9~\AA,
give values of line ratios which imply electron densities far lower (by up to an
order of magnitude)
than those indicated by other diagnostics,
\item
the observed ratios for the intensity of the 243.8~\AA\ line against 
those of the 
3s3p $^{3}$P$_{J}$--3s3d $^{3}$D$_{J^\prime}$ transitions all lie 
well outside the range of values allowed by theory.
\end{itemize}

Dufton et al. (1990) suggested that these discrepancies were due to 
either (i) errors in the adopted atomic data, arising for example from the neglect of 
fine-structure in the excitation cross section calculations (which were generated in
LS-coupling), (ii) the limited number of fine-structure levels (14) included in the
diagnostic calculations, as for high electron density flares excitation to 
higher-lying levels followed by cascades may be important,
(iii) the possibility of additional atomic processes being responsible for the
\ion{Fe}{xv} line emission, such as fluorescence. However, more recently 
Kastner \& Bhatia (2001) noted that blending in the observational data may be responsible,
although they did not investigate this possibility.

The present theoretical line ratios for \ion{Fe}{xv} discussed in Section 3 have been 
calculated using fully relativistic electron excitation rates, as opposed to those
generated in LS-coupling employed by Dufton et al. (1990).
Additionally, the ratios have been produced with a more extensive model ion
than that considered by Dufton et al., with 53 levels as opposed to 14.
Nevertheless, as noted in Section 3, the present theoretical ratios are quite 
similar to those of Dufton et al. Crucially, they do not resolve the discrepancies 
between theory and observation found for the {\em Skylab}
flare spectra, indicating that problems with the theoretical ratios are 
unlikely to be the source of the problem.

We therefore consider blending in the observations. Firstly, Dere (1978) point out that
the 233.9~\AA\ line of \ion{Fe}{xv} is partially blended with a
\ion{Ni}{xviii} transition. However, Dufton et al. (1990) note that the \ion{Ni}{xviii}
blend can be allowed for, while Keenan et al. (1997) showed that the \ion{Fe}{xv}
and \ion{Ni}{xviii} features can be well resolved and reliable line intensities 
determined. To investigate this, we have generated 
a synthetic solar flare spectrum  using the Spectral Synthesis Package 
option (ch$\underline{~~}$ss) within
Version 4.2
of the {\sc chianti} database. Details of {\sc chianti}
and instructions on its use may be found in Dere et al. (1997) and
Young et al. (2003), and also on the website
http://wwwsolar.nrl.navy.mil/chianti.html.
In our {\sc chianti}
calculations, we adopted a constant electron density of 10$^{11}$ cm$^{-3}$, the
ionization balance of Mazzotta et al. (1998), and the solar flare differential emission
measure (DEM) option. However we note that varying these parameters, even by relatively
large 
amounts, does not 
significantly affect
the following discussions nor the conclusions drawn as to the identities of
blending species.
An example of this is provided below in our analysis of the \ion{Fe}{xv}
227.2~\AA\ transition. In addition, we stress that we
are not attempting to accurately assess the amount of blending 
in the \ion{Fe}{xv} lines. Given the uncertainties in both the 
observed and theoretical \ion{Fe}{xv}
line ratios, plus possible 
errors in the calculated {\sc chianti}
flare spectrum, this would be very difficult. Instead, we are simply investigating
if blending species can be identified, and if the predicted (approximate)
level of blending is
consistent with what is observed, hence providing an explanation for the 
discrepancies found between theory and experiment for the \ion{Fe}{xv}
flare spectra.

The synthetic {\sc chianti} flare spectrum 
indicates 
that no other feature blends with the 233.9~\AA\ transition at greater than the 1\%\ 
level, so that the line intensity should be well determined, although of course
unidentified features could be present.
Unfortunately, the same argument
does not apply to the other \ion{Fe}{xv} transitions considered by 
Dufton et al. (1990). The 243.8~\AA\ line is badly blended with an \ion{Ar}{xiv}
feature (Dere 1978), which has been confirmed by Keenan et al. (2003) in their
analysis of \ion{Ar}{xiv} line ratios from the S082A flare 
spectra. Indeed, Keenan et al.
have shown that \ion{Ar}{xiv} contributes typically around 
40\%\ to the \ion{Fe}{xv}/\ion{Ar}{xiv}
243.8~\AA\ blend. Similarly, the 3s3p $^{3}$P$_{0}$--3s3d $^{3}$D$_{1}$
transition of \ion{Fe}{xv} at 224.8~\AA\ is actually listed as a \ion{S}{ix} feature 
by Dere, and the {\sc chianti}
synthetic flare spectrum indicates that \ion{S}{ix} is responsible for 
approximately 75\%\
of the measured line intensity. 

In the case of the 3s3p $^{3}$P$_{1}$--3s3d $^{3}$D$_{2}$
line of \ion{Fe}{xv} at 227.2~\AA, {\sc chianti} 
lists a \ion{C}{v} transition which is predicted to contribute 
about 10\%\ to the total measured intensity. 
Although relatively small, this degree of blending is generally
sufficient to explain
the discrepancies between theory and observation for the R = I(227.2~\AA)/I(233.9~\AA)
ratio, due primarily to the fact that R is not very sensitive to variations in
the electron density and hence small changes in the measured ratio lead to 
large ones in the derived value of N$_{\mathrm{e}}$. For example, for the 1973 August 9 flare,
Dufton et al. measured R = 0.30 which implied N$_{\mathrm{e}}$ = 10$^{9.6}$~cm$^{-3}$, 
much lower than that derived from line ratios in \ion{Fe}{xiv}
(N$_{\mathrm{e}}$ = 10$^{10.9}$~cm$^{-3}$; Keenan et al. 1991a), which is formed at a similar 
electron temperature to \ion{Fe}{xv}.
However if one assumes that the {\sc chianti}
prediction for the strength of the \ion{C}{v} blending line is correct, and hence reduce the
measured R ratio by 10\%\ so that R$_{\mathrm{deblended}}$ = 0.27, the present \ion{Fe}{xv}
line ratio calculations then indicate N$_{\mathrm{e}}$ = 10$^{10.3}$~cm$^{-3}$, 
much closer to the \ion{Fe}{xiv} value. Indeed, the \ion{C}{v} contribution to the
227.2~\AA\ feature would only need to be increased to 17\%\ for the \ion{Fe}{xv}
electron density to be the same as that deduced from \ion{Fe}{xiv}. 
We note that there are numerous uncertainties in the 
{\sc chianti} synthetic spectrum, including for example
errors in the adopted electron impact excitation rates or emission
measure distribution, and possible variations in element abundances.
Hence an increase in the 
\ion{C}{v} contribution from 10\%\ to 17\%\ would not be unreasonable. 
For example, if we reduce the adopted electron density in the {\sc chianti}
synthetic spectrum to 10$^{10}$ cm$^{-3}$, and change
the DEM from that for a flare
to the quiet Sun, the predicted \ion{C}{v}
contribution to the 227.2~\AA\ blend 
increases from 10\%\ to 50\%. This also illustrates our point above, namely that
varying
the adopted {\sc chianti} parameters by even large amounts does not affect our conclusions
regarding the identifications of line blends.

Similarly, for the 3s3p $^{3}$P$_{2}$--3s3d $^{3}$D$_{2}$ line of \ion{Fe}{xv} at 
234.8~\AA, the {\sc chianti}
synthetic spectrum indicates blending with two \ion{Ne}{iv} lines, with the 
\ion{Fe}{xv} component only contributing about 75\%\ to the measured 234.8~\AA\
intensity. Reducing the experimental I(234.8~\AA)/I(233.9~\AA)
intensity ratios to allow for this blending once again 
improves agreement between theory and observation.
For example, Dufton et al. (1990) measured the I(234.8~\AA)/I(233.9~\AA)
ratio for the 1974 January 21 flare to be 0.10, implying 
N$_{\mathrm{e}}$ = 10$^{9.5}$~cm$^{-3}$. However removing the predicted \ion{Ne}{iv}
contribution to the 234.8~\AA\ intensity reduces the experimental ratio to 0.08,
which then yields N$_{\mathrm{e}}$ = 10$^{10.5}$~cm$^{-3}$. This is 
in very good agreement 
with the value of N$_{\mathrm{e}}$ = 10$^{10.7}$~cm$^{-3}$ determined from
\ion{Fe}{xiv} (Keenan et al. 1991a).

In the case of the 3s3p $^{3}$P$_{1}$--3s3d $^{3}$D$_{1}$ feature of \ion{Fe}{xv}
at 227.7~\AA, the {\sc chianti} synthetic spectrum does not predict any 
significant blending species, with the strongest line (\ion{Fe}{xii} 227.66~\AA)
only contributing about 3\%\ to the total 227.7~\AA\ intensity.
However the discrepancy between theory and observation indicates the presence of 
a major blend. For example, in the 1973 December 17 flare, Dufton et al.
(1990) measure R = I(227.7~\AA)/I(233.9~\AA) = 0.18, which gives N$_{\mathrm{e}}$ = 
10$^{9.3}$~cm$^{-3}$,
compared to the \ion{Fe}{xiv} density estimate of N$_{\mathrm{e}}$ = 10$^{10.4}$~cm$^{-3}$
(Keenan et al. 1991a). The experimental R ratio needs to be reduced to 0.11 
in order for the \ion{Fe}{xv} density to match that deduced from \ion{Fe}{xiv},
implying that the blending species in the 227.7~\AA\ feature contributes
about 40\%\ to the total intensity.
An inspection of line lists, such as the NIST Database 
(http:$/$$/$physics.nist.gov$/$PhysRefData$/$), reveals 
the only likely candidates for the blend to be the 
2p$^{2}$ $^{3}$P$_{1}$--2p3s $^{3}$P$_{0}$
(227.63~\AA) and 2p$^{2}$ $^{3}$P$_{2}$--2p3s $^{3}$P$_{1}$ (227.69~\AA)
transitions of \ion{O}{v}. Other $\Delta$n = 1 lines of \ion{O}{v}
have been detected in the S082A flare spectra between $\sim$193--249~\AA,
arising from 2s2p--2s3s and 2s2p--2s3d transitions (Keenan et al. 1991b). 
Unfortunately, no atomic data exist for the 2p$^{2}$ $^{3}$P--2p3s $^{3}$P
lines, and hence theoretical intensity ratios cannot be generated involving these
features, to compare with the observations.
However, Huang et al. (1988) 
have detected the 
2p$^{2}$ $^{3}$P--2p3s $^{3}$P lines in a tokamak spectrum, and found their intensities
to be comparable to those of the 2s2p--2s3s and 2s2p--2s3d features. 
We are therefore confident that the blending in the 227.7~\AA\ line arises from
\ion{O}{v}.

Based on the above discussions, we conclude that the \ion{Fe}{xv}
lines considered by Dufton et al. (1990) are blended in the 
S082A flare spectra, hence providing an explanation for the 
discrepancies between theory and observation found 
by these authors. As a result, it is clear that the \ion{Fe}{xv}
features in the short wavelength S082A plates cannot be employed
as electron density diagnostics for solar flares. However, does the same argument
apply to the \ion{Fe}{xv} lines detected on the long wavelength S082A
plates? An inspection of Table 1 reveals that all of the observed line ratios 
are much larger than allowed by theory. For example, in the 1973 August 9 flare, 
the measured intensity ratio I(321.8~\AA)/I(417.3~\AA) = 1.0, while from Fig. 3
the theoretical high density limit is 0.14. Similarly the experimental
I(481.5~\AA)/I(417.3~\AA) ratio for this flare is 0.26, compared to the 
theoretical high density limit of 0.13. The {\sc chianti}
synthetic flare spectrum indicates that the 417.3~\AA\ transition should be free 
from blends, with no nearby features predicted to contribute more than about 0.5\%\
to the total measured line intensity. However, unfortunately the 
481.5~\AA\ feature is very badly blended with \ion{Ne}{v} transitions, which according to
{\sc chianti}
are responsible for around 80\%\ of the total line flux. 

In the case of the 321.8 and 327.0~\AA\ lines of \ion{Fe}{xv},
{\sc chianti} indicates either the presence of no blending species (327.0~\AA),
or only a small ($\sim$20\%) contribution from \ion{Fe}{x} transitions
(321.8~\AA). A search of line lists also reveals no likely blends.
Furthermore, intensity ratios involving these lines in active region spectra
show good agreement between theory and observation (Keenan et al. 2005), and one
would actually expect \ion{Fe}{x} to make a smaller contribution to 
the 321.8~\AA\ line intensity in a flare than in an active region, given that 
\ion{Fe}{xv} is more highly ionized and hence should be relatively much stronger
in the former. 
(Indeed, {\sc chianti} predicts that \ion{Fe}{x} contributes 
around 50\%\ of the 321.8~\AA\ line intensity in an active region).
Given these facts, it is difficult to believe that blending is responsible 
for the discrepancies between theory and observation for the I(321.8~\AA)/I(417.3~\AA)
and I(327.0~\AA)/I(417.3~\AA) intensity ratios. Significant problems with the measurements of 
the 417.3~\AA\ line can also be ruled out, as the experimental
I(321.8~\AA)/I(327.0~\AA)
ratios similarly show discrepancies with theory, being up to
$\sim$75\%\ larger than the theoretical high density limit of 0.55. 
We therefore believe that the discrepancies are due to intensity calibration 
uncertainties 
in the S082A observations, most probably arising from the fact that 
the 321.8 and 327.0~\AA\ lines lie close to the short wavelength cutoff
of the plates, and hence are more susceptible to abberation effects 
(see Section 2).
Support for this comes from a spectrum of a solar flare in the 280--330~\AA\
region (Zhitnik et al. 2005), obtained by the RES--C 
spectroheliograph on the CORONAS--F orbital station, at a resolution
of 0.1~\AA.
Although most of the \ion{Fe}{xv} lines in the RES--C flare spectrum
are blended (as in the S082A data), a reliable measurement is available for
the \ion{Fe}{xv} intensity ratio I(321.8~\AA)/I(327.0~\AA) = 0.48$\pm$0.05. This 
implies N$_{e}$ = 10$^{10.5\pm0.4}$~cm$^{-3}$ (Keenan et al. 2005), 
in excellent agreement with typical flare densities derived from, for example,
\ion{Fe}{xiv} line ratios (Keenan et al. 1991a).

In Table 2 we summarise our identifications of lines which blend with the 
\ion{Fe}{xv}
transitions in the S082A flare spectra. 
On the basis of this, we may confidently state that
most of the \ion{Fe}{xv}
lines in solar flare spectra are blended (at least at spectral resolutions of
$\sim$0.1~\AA). There is hence no need to invoke errors in atomic data or other
reasons to explain discrepancies between theory and observation, as suggested by
Dufton et al. (1990). As a consequence, \ion{Fe}{xv} flare lines do not provide 
useful electron density diagnostics for the emitting plasma, with the exception 
of the I(321.8~\AA)/I(327.0~\AA) intensity ratio.
 
%
%
%
\begin{table}
\caption{Summary of blending species for \ion{Fe}{xv} transitions in S082A flare spectra}       
\label{table:1}      
\begin{tabular}{l l}      
\hline               
\ion{Fe}{xv} line (\AA) & Possible blending species 
\\  
\hline    
224.8 & Mostly due to \ion{S}{ix} 224.8~\AA 
\\ 
227.2 & \ion{C}{v} 227.2~\AA\ contributes 10--20\%\ to the blend
\\
227.7 & Major blend with \ion{O}{v} 227.6 and 227.7~\AA
\\
233.9  
\\
234.8 & \ion{Ne}{iv} 234.7~\AA\ contributes about 20\%\ to the blend 
\\
243.8 & Badly blended with \ion{Ar}{xiv} 243.8~\AA
\\
321.8 & \ion{Fe}{x} 321.8~\AA\ contributes about 20\%\ to the blend
\\
327.0  
\\
417.3 
\\
481.5 & Badly blended with \ion{Ne}{v} 481.4~\AA
\\
\hline
\end{tabular}

\end{table}

\begin{acknowledgements}
K.M.A. and D.S.B. acknowledge financial support from EPSRC and PPARC, respectively.
F.P.K. is grateful to AWE Aldermaston for the award of a William Penney
Fellowship. {\sc chianti} is a collaborative project 
involving the Naval Research Laboratory (USA), Rutherford
Appleton Laboratory (UK), and the Universities of Florence 
(Italy) and Cambridge (UK).
\end{acknowledgements}

\end{document}